\documentclass{webofc}
\usepackage[utf8]{inputenc}
\usepackage[varg]{txfonts}   
\usepackage{latexsym}
\usepackage{amssymb}
\usepackage{amsmath}
\usepackage{amstext}
\usepackage{graphics}
\usepackage{float}
\begin{document}
\title{Discovery of the Odderon by TOTEM experiments and the FMO approach}
%
%

\author{\firstname{Evgenij} \lastname{Martynov}\inst{1,3}\fnsep\thanks{\email{martynov@bitp.kiev.ua}} \and
        \firstname{Basarab} \lastname{Nicolescu}\inst{2}\fnsep\thanks{\email{basarab.nicolescu@gmail.com}} 
}

\institute{Bogolyubov Institute for Theoretical Physics, Metrologichna 14b, Kiev, 03680 Ukraine 
\and
           Faculty of European Studies, Babes-Bolyai University, Emmanuel de Martonne Street 1, 400090 Cluj-Napoca, Romania 
 }

\abstract{%
This paper is an extended version of the talk by B. Nicolescu at the XLVIII International Symposium on Multiparticle Dynamics (ISMD2018) at Singapore, 3-7 September, 2018. Theoretical basis and history of the Froissaron and Maximal Odderon (FMO) approach for elastic $pp$ and $\bar pp$ scattering is presented. Precise formulation of the FMO model at any momentum transfer squared $t$ is given. The model is applied to description and analysis of the experimental data in a wide interval of energy $\sqrt{s}$ and  $t$.  The special attention is given for the latest TOTEM data at 13 TeV, both at $t=0$ and at $t\neq 0$ and to their interpretation in the FMO model. It is emphasized that the last TOTEM results can be considered as clear evidence for the first experimental observation of the Odderon, predicted theoretically about 50 years ago.
}
\maketitle
\section{Introduction}
\label{intro}
Recently, the TOTEM experiment released the following values at $\sqrt{s}=13$ TeV of $pp$ total cross section $\sigma^{pp}$ and 	$\rho^{pp}$ parameter \cite{TOTEM-1, TOTEM-2}
\begin{equation}\label{eq:exp-sigma-rho}
\sigma^{pp}=110\pm 6 \quad \text{mb}, \qquad \rho^{pp}=0.098\pm 0.01\quad ( 0.10 \pm  0.01)
\end{equation}

The value of $\sigma^{pp}$ is in good agreement with the standard best COMPETE prediction \cite{COMPETE}  but is in violent disagreement with the COMPETE prediction for  $\rho^{pp}$ (which is much higher than the experimental value). This is the first problem we have to solve before drawing conclusions about the discovery of the Odderon (which is absent in the COMPETE approach). On the other hand, the experimental value of $\rho^{pp}$ is in perfect agreement with the Avila–Gauron–Nicolescu (AGN) model \cite{AGN}, which includes the Odderon and which predicts a value of 0.105. In fact, the AGN model was the only  model which correctly predicted $\rho^{pp}$ but it predicted also higher values of $\sigma^{pp}$ than the TOTEM values, a discrepancy which might be connected with the ambiguities in continuing the amplitudes in the non-forward region. This is the second problem we have to solve before drawing conclusions about the discovery of the Odderon. The third task is to extend the new FMO, model which fixes the first two problems, for $t\neq 0$ and to compare it with the newest TOTEM data on differential cross section at $\sqrt{s}=13$ TeV. 

Part of this program was realized in the Refs. \cite{MN-0, MN-1, MN-2}. Here we give only the main, principal results obtained in this direction. Many additional results and details can be found in the refs. \cite{MN-0, MN-1, MN-2}.  

\section{Theoretical justification and history of the FMO approach}
\label{sec-1}
The aforementioned AGN model is not the only realization of Froissaron and Maximal Odderon (FMO) approach to high energy elastic hadron scattering. The general principles and basic strict results of Quantum Field Theory and Analytic $S$-matrix theory can not determine in an unique way the amplitudes  of elastic scattering. At the same time they  (many of them as well as references of original papers, where they have been obtained,can be found in review \cite{Eden}) strongly restrict the arbitrariness in construction of the models. Additional restrictions for the model are given by available experimental data. 

In this Section we discuss  the theoretical constraints which should be satisfied in any model of elastic scattering and some additional assumptions which lead to FMO-type models. Omitting the details of the strict results and assumptions we rather give the list of those that are important in Froissaron anf Maximal Odderon approach to high energy elastic scattering of hadrons (moreover we concentrate on $pp$ and $\bar pp$ elastic scattering.) As usually, as it is made for processes at high energy, we consider proton and antiproton as spin-less particles. 

\subsection{Rigorous results  and assumptions important for FMO model}

{\bf Froissart-Martin-Lukaszuk bound}.
It follows from unitarity and analyticity of amplitude that 
\begin{equation}\label{eq:Froissart bound} 
\sigma_t(s)\leq C\ln^2(s/s_0), \qquad C\leq  \dfrac{\pi}{m_\pi^2}.
\end{equation}
Here and in what follows we take $s_0=1$ Gev.

In Regge theory the contribution of Regge pole with trajectory $\alpha(0)$ to the elastic scattering amplitude has at high $s$ the form
\begin{equation}\label{eq:Regge pole}
F(s,t)\propto (-is/s_0)^{\alpha(t)}.
\end{equation}
From the Eq. (\ref{eq:Froissart bound}) one can obtain the unitarity bound for the intercept of trajectory 
\begin{equation}\label{eq:intercept bound}
\sigma_t(s)\approx \dfrac{1}{s}\text{Im}F(s,0)\propto (s/s_0)^{\alpha(0)-1}\qquad \Rightarrow \qquad \alpha(0)\leq 1.
\end{equation} 
{\bf Pomeranchuk theorems}. 
	\begin{equation}\label{eq; Pomeranchuk-1}
	\begin{array}{lll}
\text{1. If} \quad &\sigma_t^{ab,\bar ab}(s)\to const \quad &\text{at} \quad s\to \infty \quad \text{then} \quad	\sigma_t^{ab}(s)-\sigma_t^{\bar ab}(s)\to 0\\
\text{2. If}\quad &\sigma_t^{ab,\bar ab}(s)\to \infty \quad  &\text{at} \quad s\to \infty \quad \text{then}\quad  	\sigma_t^{ab}(s)/\sigma_t^{\bar ab}(s)\to 1
\end{array}
	\end{equation}
{\bf Eden theorem}.
In accordance with crossing symmetry we consider two terms, crossing-even (CE) and crossing-odd (CO), of amplitudes. 
\begin{equation}\label{eq:pp-amplitude}
F^{pp}(s,t)\equiv F^{pp}(z_t,t)=F_+(z_t,t)+F_-(z_t,t),
\end{equation}
\begin{equation}\label{eq:pap-amplitude}
F^{\bar pp}(s,t)\equiv F^{\bar pp}(z_t,t)=F_+(z_t,t)-F_-(z_t,t)
\end{equation}
\begin{equation}\label{eq:CE-CO-amplitudes}
F_\pm(\pm z_t,t)=\pm F_\pm(z_t,t)
\end{equation}
where $-z_t=\cos\theta_t=1+2s/(t-4m^2)$.

Generally, the Eden theorem claims that\\ if at $s\to \infty$
\begin{equation}\label{eq:Eden-1}
\sigma^{ab}_{\bar ab}(s)\sim C\ln^{\mu_+} (s/s_0)\pm D\ln^{\mu_- }(s/s_0)
\end{equation}
and if $\mu_+\leq 2$ then
\begin{equation}\label{eq:Eden-2}
\mu_- \leq \mu_+/2+1.
\end{equation}
Thus for difference of cross sections we have
\begin{equation}\label{eq:Eden-3}
\Delta \sigma =|\sigma^{ab}(s)-\sigma^{\bar ab}(s)|\leq 2D\ln^{\mu_+/2}(s/s_0).
\end{equation}
{\bf Cornille-Martin theorem} \cite{Cornille-Martin} is an analog of the second Pomeranchuk theorem.

If $s\to \infty$ and $|t|\leq t_0/\ln^2(s/s_0)\to 0 $ where $t_0, s_0 $ are constants  then
\begin{equation}\label{eq:Eden-4}
\dfrac{d\sigma^{\bar pp}}{dt}\big /\dfrac{d\sigma^{pp}}{dt}\to 1.
\end{equation}
{\bf Auberson-Kinoshita-Martin theorem} is very important for a construction of FMO model at $t\neq 0$.
It was proved \cite{AKM} that in the case when $\sigma_t(s)\propto \ln^2(s/s_0), \quad s_0=const $ at $s\to \infty $
\begin{equation}\label{eq::AKM theorem} 
A_\pm (s,t)/A_{\pm}(s,0)=f(\tau)  \quad \text{where} \quad  \tau=r\sqrt{-t}\ln(s/s_1). \quad s_1=const. 
\end{equation}
{\bf Dispersion relations} in an integral form can be derived from analyticity of the amplitude and Cauchy theorem. At high energy the derivative dispersion relations \cite{BKS} \cite{KN} for amplitudes are useful for construction of FMO model.  
\begin{equation}\label{eq:DDR}
\begin{array}{ll}
\text{Re} [F_+(z_t,t)/s]&=\quad \left [ \dfrac{\pi}{2}\dfrac{\partial}{\partial \xi} +\cdots \right ]\text{Im} [F_+(z_t,t)/s],\\
\dfrac{\pi}{2}\dfrac{\partial}{\partial \xi }\text{Re} [F_-(z_t,t)/s]&=-\left [1-\dfrac{1}{3}\left (\dfrac{\pi}{2}\dfrac{\partial}{\partial \xi } \right )^2 +\cdots \right ] \text{Im} [F_-(z_t,t)/s]
\end{array}
\end{equation}
{\bf Unitarity bounds for partial and impact parameters amplitude.}
There is well known unitarity equation for $s$-channel partial amplitude 
\begin{equation}
\text{Im}\,a_l(s)=\dfrac{1}{16\pi}\sqrt{1-4m^2/s}\,|a_l(s)|^2+\text{inelastic contribution}.
\end{equation}
Inelastic contribution is positive, therefore we have the important bounds for partial amplitudes at $s\ge 4m^2$
\begin{equation}\label{eq:partial bounds}
0\leq \text{Im} a_l(s) \leq |a_l(s)|^2\leq 1. 
\end{equation}
Similarly one can obtain the unitarity equation for impact parameter amplitude 
\begin{equation}\label{eq:impact-bound-1} 
\text{Im}H(s,b)=|H(s,b)|^2+G_{inel}(s,b), \qquad H(s,b)=\dfrac{1}{8\pi  s}\int\limits_0^\infty dqqJ_0(bq)F(s,-q^2).
\end{equation}
with the bounds
\begin{equation}\label{eq:impact-bound-2} 
0\leq \text{Im}H(s,b) \leq |H(s,b)|^2\leq 1. 
\end{equation}
{\bf Maximality principle for strong interactions} was formulated for the first time in 1962 by G. Chew \cite{Chew} for simple Regge poles. Later it was reformulated as the following: 

''The cross sections of hadron interactions at high energies should saturate the asymptotic bounds in their functional form.`` 

This principle was taken as a ground for construction of the models of Froissaron (which can be named as Maximal Pomeron) and of Maximal  Odderon. In fact, Froissaron realizes the maximality principle for total cross sections  while the Maximal Odderon makes the same for difference of $|\sigma_t^{\bar pp}-\sigma_t^{pp}|$.  Thus for elastic $pp$ and $\bar pp$ scattering the FMO model in its maximal variant   is a model in which
\begin{equation}\label{eq;FMO general}
\sigma_t^{pp, \bar pp}(s) \propto \ln^2(s/s_0) \qquad \text{and} \qquad
|\sigma_t^{pp}(s)-\sigma_t^{\bar pp}(s)| \propto \ln(s/s_0)
\end{equation} 

\subsection{Short history of FMO approach}
The first realizaion of the maximality in strong interactions was considered in 1973   by L. {\L}ukaszuk and B. Nicolescu \cite{LN}.

In 1975 K. Kang and B. Nicolescu \cite{KN} considered in detail the model at $t=0$ comparing it with the experimental data available at the time.  

The name {\bf Odderon} for the first time was proposed by D. Joynson, E. Leader, B. Nicolescu and C. Lopez in \cite {JLNL} two years after the first paper of {\L}ukaszuk and B. Nicolescu  .

The detailed investigation of unitarity and analyticity of Maximal Odderon was performed by P. Gauron, {\L}ukaszuk and B. Nicolescu \cite{GLN}. It was shown in this paper that FMO model does not contradicts the main theorems and bounds on amplitudes and observable quantities obtained in $S$-matrix theory.  

Then FMO model was developed and improved in many papers, the last of them in particular AGN model (R. Avila, P. Gauron and B. Nicolescu)  \cite{AGN} and its minor modification \cite{MN-AGN} were published in 2007-2008.    The low values of $\rho_{pp}$ at LHC (but with too high value of $\sigma_t$)  were predicted \cite{AGN}.
The alternative for AGN model  \cite{EM} was suggested  in 2007. In this model another form for the  Froissaron and Maximal Odderon terms had been considered.

New stage in developing the FMO approach is caused by the TOTEM measurement and results obtained at 13 TeV.

\section{Formulation of the FMO model at any $t$}
\label{sec-2}
Our aim to construct the FMO model for $pp$ and $\bar pp$ amplitudes and then to compare the model with the experimental data on $\sigma_t(s)$, $\rho(s)$ and $d\sigma(s)/dt$ which are related with amplitudes as 
\begin{equation}\label{eq: observ}
\begin{array}{ll}
\sigma_t(s)&=\dfrac{1}{\sqrt{s (s-4m^2 )}}\text{Im} F(s,0), \\
\dfrac{d\sigma_{el}}{dt}&=\dfrac{1}{64\pi ks(s-4m^2)}|F(s,t)|^2
\end{array}
\end{equation}
where $k=0.3893797\,\, \text{mb}\cdot\text{GeV}^2$. With this normalization the amplitudes have dimension $\text{mb}\cdot\text{GeV}^2$.  

The amplitudes of proton-proton and antiproton-proton scattering are defined by Eqs. (\ref{eq:pp-amplitude})-(\ref{eq:CE-CO-amplitudes})
In the FMO model CE (crossing-even) and CO (crossing-odd) terms of amplitudes are defined as sums of the asymptotic contributions $F^H(s,t)$,  $F^{MO}(s,t)$  and Regge pole contributions which are important at the intermediate and relatively low energies
\begin{equation}\label{eq:FMO+-R}
\begin{array}{l}
F_+(z_t,t)=F^H(z_t,t)+F^{R_+}(z_t,t), \\
F_-(z_t,t)=F^{MO}(z_t,t)+F^{R_-}(z_t,t)
\end{array}
\end{equation} 
where $F^H(z_t,t)$ denotes the Froissaron contribution and $F^{MO}(z_t,t)$ denotes the Maximal Odderon contribution. $F^{R_\pm}(z_t,t)$ stand for the standard Pomeron, Odderon, secondary reggeons and their double exchanges. Their specific form will be defined in the next two subsections.

\subsection{Froissaron and Maximal Odderon contributions in FMO amplitudes}\label{sec:Gen.Con}
We assume that in FMO model at $s\to \infty $
\begin{equation}\label{eq:FMO-0}
\sigma_t(s)\propto \ln^2(s/s_0), \qquad \Delta \sigma(s)=|\sigma_t^{pp}-\sigma_t^{\bar pp}|\propto \ln(s/s_0)
\end{equation} 
It means that we consider $\mu_+=2$ and $\mu_-=1$ in the Eqs. (\ref{eq:Eden-1},\ref{eq:Eden-2}).
One can show from the AKM theorem (\ref{eq::AKM theorem}) or making use the simple arguments from \cite{EM} that the partial amplitude for the main Froissaron term at $\omega=j-1\approx 0$ has the form
\begin{equation}\label{eq:phi-F-MO}
\varphi_{\pm}(\omega,t)=\binom{i}{1}\dfrac{\beta_{\pm} (\omega,t)}{\left
	[\omega^2+\omega^2_{\pm}\right]^{3/2}}, \qquad \omega_{\pm}=r_{\pm}\sqrt{-t/t_0}, \quad t_0=1\text{GeV}^2.
\end{equation}

Constructing the FMO model we assume (it is only one of the possibilities, another alternatives would be considered in further investigations) that in accordance with a structure of the singularity of $\varphi_\pm(\omega,t)$ at $\omega^2+\omega_{0\pm}^2=0$  ($\omega_{0\pm}^2= R_\pm^2q_{\perp}^2)$ the functions $\beta_\pm(\omega,t)$ depend on $\omega$ through the variable $\kappa_\pm =(\omega^2+\omega_{0\pm}^2) ^{1/2}$. Then they can be expanded in powers of $\kappa_\pm$.
\begin{equation}\label{eq:expand2}
\beta_\pm(\omega,t)=\beta_{\pm,1}(t)+(\omega^2+\omega_0^2)^{1/2}  \beta_{\pm,2}(t)+(\omega^2+\omega_0^2)\beta_{\pm,3}(t).
\end{equation}

Thus we have the following CE and CO amplitudes in the FMO model
\begin{equation}\label{eq:F-MN}
\begin{array}{rl}
\dfrac{-i}{z}F^H(z_t,t)&=H_1\zeta^2\dfrac{2J_{1}(r_{+}\tau \zeta)} {r_{+}\tau \zeta }\Phi^{2}_{H,1}(t)
+H_2\zeta\dfrac{\sin(r_{+}\tau \zeta)}{r_+\tau \zeta}\Phi^{2}_{H,2}(t)
+(H_3-C^P)\Phi^{2}_{H,3}(t),\\
\Phi_{H,i}(t)&=\exp(b^H_iq_+),\quad i=1.2, \\
\Phi_{H,3}(t)&=h\exp(b^H_3q_+)+(1-h)\exp(b^H_4q_+), \\
q_+&=2m_{\pi }-\sqrt{4m_{\pi }^2-t}.\\   
\end{array}
\end{equation}

\medskip

\begin{equation}\label{eq:MO-MN}
\begin{array}{rl}
\dfrac{1}{z}F^{MO}(z_t,t)&=O_1\zeta^2\dfrac{2J_{1}(r_{-}\tau \zeta)}{r_{-}\tau \zeta}\Phi^{2}_{O,1}(t) 
+ O_2\zeta \dfrac{\sin(r_{-}\tau \zeta)}{r_{-}\tau \zeta}\Phi^{2}_{O,2}(t) 
+(O_3+C^O)\Phi^{2}_{O,3}(t),\\
\Phi_{O,i}(t)&=\exp(b^O_iq_-), \quad i=1,2,\\
\Phi_{O,3}(t)&=o\exp(b^O_3q_-)+(1-o)\exp(b^O_4q_-),\\
q_-&=3m_{\pi }-\sqrt{9m_{\pi }^2-t}.\\   
\end{array}
\end{equation}

The third terms in Eq. (\ref{eq:expand2}) mimic the contribution of simple Regge poles with intercepts one. Comparing the model with data we have found that the best description is achieved when these terms are chosen in a simplified form as is given in Eqs. (\ref{eq:F-MN},\ref{eq:MO-MN}).   
\subsection{Standard Pomeron and Odderon, PP, PO and OO cuts, secondary reggeons in FMO model}

The full form of the FMO amplitudes is defined as follows
\begin{equation}\label{eq:FMO+R}
\begin{array}{l}
F_+(z_t,t)=F^H(z_t,t)+F^P(z_t,t)+F^{R_+}(z_t,t)+F^{PP}(z_t,t)+F^{OO}(z_t,t), \\
F_-(z_t,t)=F^{MO}(z_t,t)+F^{R_-}(z_t,t)+F^{PO}(z_t,t)
\end{array}
\end{equation} 
where $F^H(z_t,t)$ denotes the Froissaron contribution and $F^{MO}(z_t,t)$ denotes the Maximal Odderon contribution. and where
 $F^P(z_t,t), F^O(z_t,t)$ are simple $j$-pole Pomeron and Odderon contributions and $F^{R_+}(z_t,t),  F^{R_-}(z_t,t)$ are effective $f$ and $\omega$ simple $j$-pole ($j$ is an angular momenta of these reggeons) contributions. $F^{PP}(z_t,t)$,  $F^{OO}(z_t,t),$ $F^{PO}(z_t,t),$ are double $PP, OO, PO$ cuts, correspondingly. We consider the model at $t\neq 0$ and at energy $\sqrt{s}> 19$ GeV, so we neglect the rescatterings  of secondary reggeons with $P$ and $O$. In the considered kinematical region they are small. Besides, because $f$ and $\omega$ are effective, they  can take into account small effects from the cuts. The standard Regge pole contributions have the form
\begin{equation}\label{eq:sec Regge-st}
F^{{\cal R}_\pm}(z_t,t)=-\binom{1}{i}
2m^2C^{{\cal R}_\pm}(t)(-iz_t)^{\alpha_\pm(t)}
\end{equation}
where ${\cal R}_\pm= P,O,R_+,R_-$ and $\alpha_P(0)=\alpha_O(0)=1$. The factor $2m^2=z_t/z_t(t=0)$ (at $s\gg m^2$) is inserted in amplitudes  $F^{R_\pm}(z_t,t)$ in order to have the normalization for amplitudes and dimension of coupling constants (in mb) coinciding with those in the ref. \cite{MN-0}.

For  secondary reggeons we have considered the functions in the exponential form $C^{R_\pm}(t)=C^{R_\pm}e^{2b_{R_\pm}t}$ while for $P$ and $O$ they are chosen in a more general form
\begin{equation}\label{eq:nonexp-coupling}
\begin{array}{ll}
C^{P,O}(t)&=C^{P,O}\left [\Psi^{P,O}(t) \right ]^2, \\
\Psi^{P,O}(t)&=c^{P,O}e^{b_1^{P,O}t}+(1-c^{P,O})e^{b_2^{P,O}t}.
\end{array}
\end{equation}
which allow to take  into account some possible effects of non-exponential behavior of coupling function. 

The  double cuts are written in a simplified form as compared with their exact form. Because of free parameters $b$ in the exponents they can be considered also as effective $PP, OO, PO$ cuts. Namely,
\begin{equation}\label{eq:PP-PO-OO}
\begin{array}{ll}
F^{PP}(z_t,t)&=-\dfrac{2m^2C^{PP}}{\ln(-iz_t)}\left (-iz_t \right )^{\alpha_{PP}(t)}e^{2b^{PP}t}, \quad
\alpha_{PP}(t)=1+ \alpha'_{PP}t, \quad  \alpha'_{PP}=\alpha'_{P}/2,\\
F^{OO}(z_t,t)&=-\dfrac{2m^2C^{OO}}{\ln(-iz_t)}\left (-iz_t \right )^{\alpha_{OO}(t)}e^{2b^{OO}t}, \quad
\alpha_{OO}(t)=1+ \alpha'_{OO}t, \quad  \alpha'_{OO}=\alpha'_{O}/2,\\
F^{PO}(z_t,t)&=i\dfrac{2m^2C^{PO}}{\ln(-iz_t)}\left (-iz_t \right )^{\alpha_{PO}(t)}e^{2b^{PO}t}, \quad 
\alpha_{PO}(0)=1+ \alpha'_{PO}t,\quad \alpha'_{PO}=\dfrac{\alpha'_{P}\alpha'_{O}}{\alpha'_{P}+\alpha'_{O}}.
\end{array}
\end{equation}

\section{Comparison of the FMO model with experimental data} 
\label{sec:4}

We give here the results of the fit to the data in the following region of $s$ and $|t|$. 
\begin{equation}\label{eq:dataregion}
\begin{array}[]{llllll}
\text{for} \quad \sigma_{tot}(s), \rho(s) \quad & \text{at} \quad  5 & \text{GeV} & \leq \sqrt{s} & \leq 13 & \text{TeV}, \\
\text{for} \quad d\sigma(s,t)/dt \quad & \text{at} \quad 9 & \text{GeV} & \leq \sqrt{s} & \leq 13 & \text{TeV}\\ 
\nonumber
\text{and}  & \text{at}  \quad 0.05  & \text{GeV}^2 & \leq |t| & \leq  5 & \text{GeV}^2.
\end{array}
\end{equation}
The $t$-region is chosen in such a way that we can  ignore the contribution of the  Coulumb part of amplitudes which in given region are one order of  magnitude or less than 1\% of the nuclear amplitude.

For 13 TeV TOTEM data we used the data at $t=0$ for  $\sigma_{tot}$ \cite {TOTEM-1} and $\rho$ \cite{TOTEM-2} and data for $d\sigma/dt$  at $t\neq 0$ presented at the 4th Elba Workshop on Forward Physics @ LHC Energy  by F.Nemes  \cite{Nemes-2018} and at the 134th open LHCC meeting by F. Ravera \cite{Ravera-2018}.

\subsection{Total $pp$ and $\bar pp$ cross sections and parameters $\rho_{pp}$ and $\rho_{pp}\bar pp$}

\subsubsection{FMO model in its maximal form at $t=0$}
\begin{figure}[H]
	\includegraphics[width=0.5\linewidth]{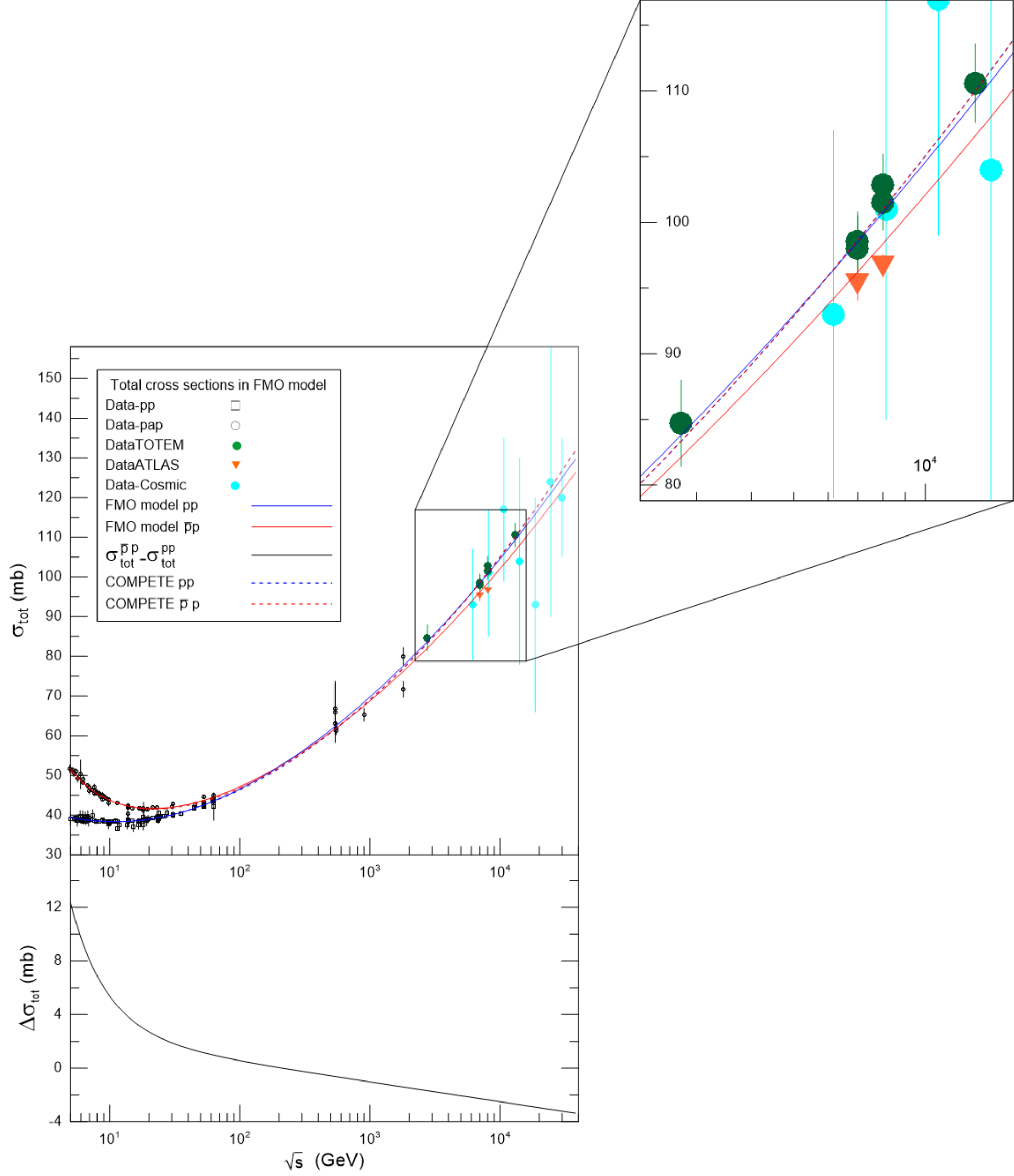}
	\includegraphics[width=0.35	\linewidth]{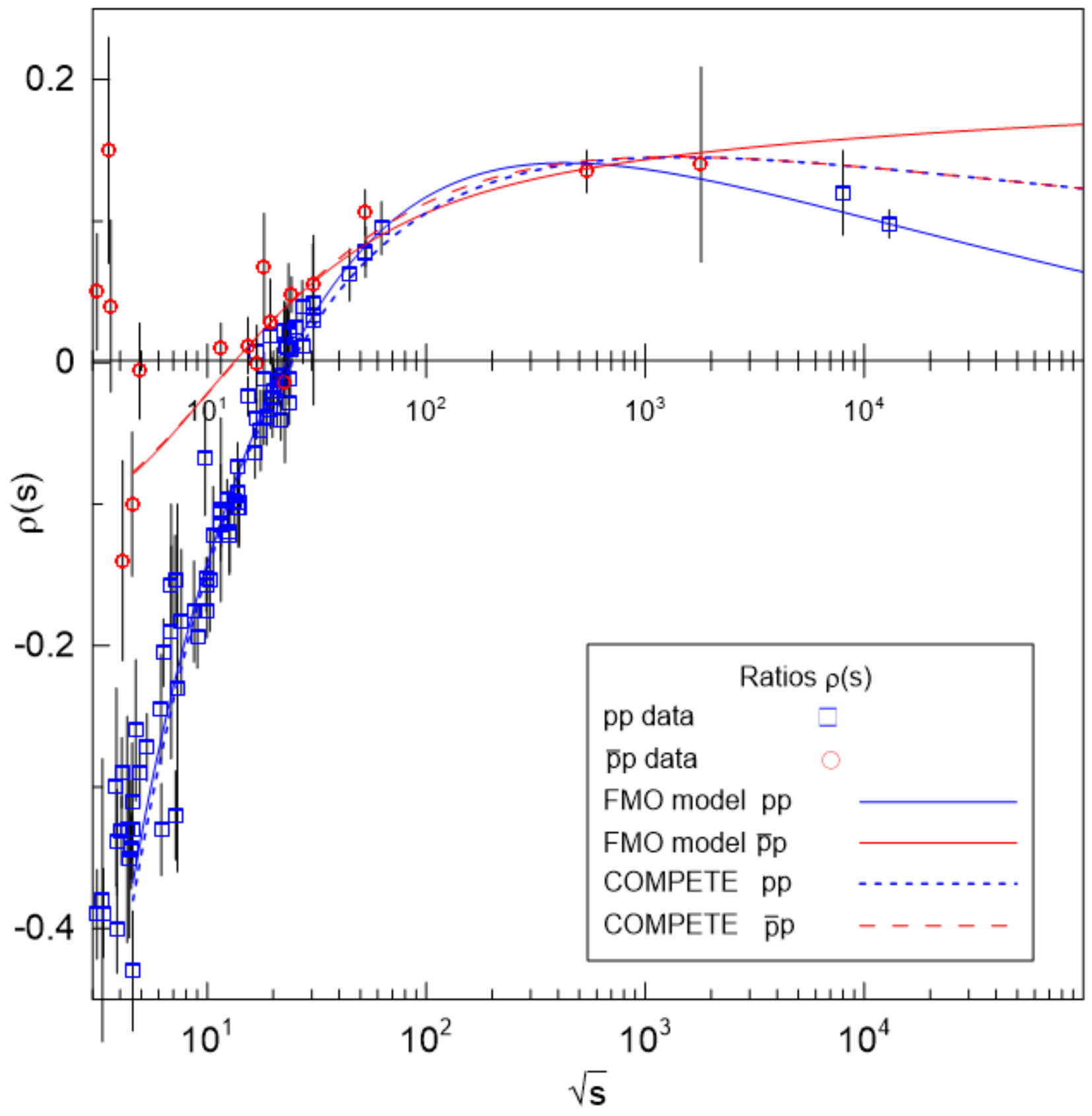}
	\caption{Total cross sections and ratios of the real to imaginary part of the forward elastic amplitude in FMO model (solid lines). The curves (dashed lines) of the best COMPETE fit \cite{COMPETE} are shown also for a comparison}
	\label{fig:fig1}
\end{figure}

In the Ref. \cite{MN-0} the $pp$ and $\bar pp$ amplitudes were defined in accordance with Eqs. (\ref{eq:pp-amplitude}, \ref{eq:pap-amplitude}) and (\ref{eq:FMO+R}) without $PP, PO, OO$ terms. 
The contributions of Froissaron and Odderon ($F^H_+(z)$ and $F^{MO}_-(z_t)$,  correspondingly) are parameterized at $t=0$ in terms of 6 parameters,  
\begin{equation}\label{eq:F-at-t=0}
\begin{array}{ll}
F^H_+(z)&=i(s-2m^2)[H_1\ln^2(-iz)+H_2\ln(-iz)+H_3], \\
F^{MO}_-(z)&=(s-2m^2)[O_1\ln^2(-iz)+O_2\ln(-iz)+O_3]
\end{array} 
\end{equation}
The standard Pomeron and Odderon in  \cite{MN-0} were included to constant terms $H_3$ and $O_3$, correspondingly, because they give constant contributions at $t=0$. The contributions of the secondary reggeons contain additionally 4 parameters
\begin{equation}\label{eq:sec Regge-0}
F^R_\pm(z)=-\binom{1}{i}
C^R_\pm (-iz)^{\alpha_\pm(0)}
\end{equation}
Quality of the fit is shown at the Table 1 and in the Fig. \ref{fig:fig1}. The values of parameters and more details are given in  \cite{MN-0}.
 \renewcommand{\arraystretch}{1.3}
\begin{table}[H]
	\centering
	\begin{tabular}{ccc}
		\hline 
		Observable& Number of points & $\chi^2/N_p$	\\ 
		\hline 
		$\sigma_{tot}^{pp}$& 110 & 0.8486	\\ 
		$\sigma_{tot}^{\bar pp}$& 59 & 0.8662\\ 
		$\rho^{pp}$& 66 & 1.6088\\ 
		$\rho^{\bar pp}$& 11 & 0.5468 \\ 
		\multicolumn{2}{c}{ $\chi^2/\text{dof}$} & 1.0871 \\
		\hline 
	\end{tabular} 
	\label{tab:tab1}
	\caption{Number of experimental points $N_p$ and $\chi^2/N_p$ for $\sigma_{tot}$ and $\rho$ in the fit with FMO model}
\end{table}
This study shows that, the new TOTEM datum $\rho_{pp}$ = 0.1$\pm$ 0.01 can be considered as
the first certain experimental discovery of the Odderon, namely in its maximal form.

\subsubsection{FMO model in a more general form at $t=0$}

It is important to check if the Froissaron-Maximal Odderon (FMO) approach is the only model in agreement with the LHC data. We put the question: is the maximality of strong interactions not only in agreement with experimental data but it is even required by them? To find the answer on the question we have considered  the generalized the FMO approach by relaxing the $\ln^2(s/s_0)$ constraints both in the even-and odd-under-crossing amplitude \cite{MN-1}. 

So, we considered the following form of the amplitudes:
\begin{equation}\label{eq:Ft=0}
F^H_+(z)=i(s-2m^2)[H_1\ln^{\mu_+}(-iz)+H_2\ln^{\mu_+-1}(-iz)+H_3], 
\end{equation}
\begin{equation}\label{eq:MOt=0}
F^{MO}_-(z)=(s-2m^2)[O_1\ln^{\mu_-}(-iz)+O_2\ln^{\mu_--1}(-iz)+O_3], 
\end{equation}
\begin{equation}\label{eq:sec Regge-4}
F^R_\pm(z)\equiv \binom{P,R_+}{O,R_-}=-\binom{1}{i}C^R_\pm (-iz)^{\alpha_\pm(0)}.
\end{equation} 
For $\mu_+=\mu_-=2$ and $\alpha_P(0)=\alpha_O(0)=1$ we get exactly the FMO model  of the ref. \cite{MN-0}.
Our aim is verify which values of   $\mu_+$ and $\mu_-$, as well as which values of  $\alpha_P(0)$ and $\alpha_O(0)$ are the best for fitting all existing experimental data on $\sigma_{tot}(s)$ and $\rho(s)$.

The parameters $\mu_+$ and $\mu_-$ are not arbitrary. They are constrained by analyticity, unitarity, crossing-symmetry and positivity of cross sections  (see Eqs. (\ref{eq:Eden-1}-\ref{eq:Eden-3})). 

The results of fitting are quite spectacular (see the details in \cite{MN-1}): the values of $\mu_+$ and $\mu_-$ come back to the saturation values $\mu_+=2$ and $\mu_-=2$. Intercepts $\alpha_P(0)$ of Pomeron when it is free comes back to the value $\alpha_P(0)=1$.  
Thus we show that, in spite of a considerable freedom of a large class of amplitudes, the experimental data choose the maximal form of the FMO model, namely  the maximal growth with energy of $\sigma_{tot}(s)$ and  the maximal growth of the difference $|\sigma_{tot}^{pp}(s)-\sigma_{tot}^{\bar pp}(s)|$ \cite{Eden}.

\section{FMO and experimental data on $d\sigma(s,t)/dt$}
\label{sec:5}
The FMO model defined by Eqs.~(\ref{eq:F-MN}-\ref{eq:PP-PO-OO}) was applied to describe simultaneously  the $pp$ and $\bar pp$ total cross sections $\sigma_t^{pp}(s),\,\, \sigma_t^{\bar pp}$, ratios $\rho^{pp}(s),\,\, \rho^{\bar pp}(s)$ and differential cross sections $d\sigma^{pp}(s,t)/dt,\,\, d\sigma^{\bar pp}(s,t)/dt$ in the region of $s$ and $t$ described in the beginning of the Section \ref{sec:4}. 

\begin{figure}[H]
	\centering
	\includegraphics[width=0.8\linewidth]{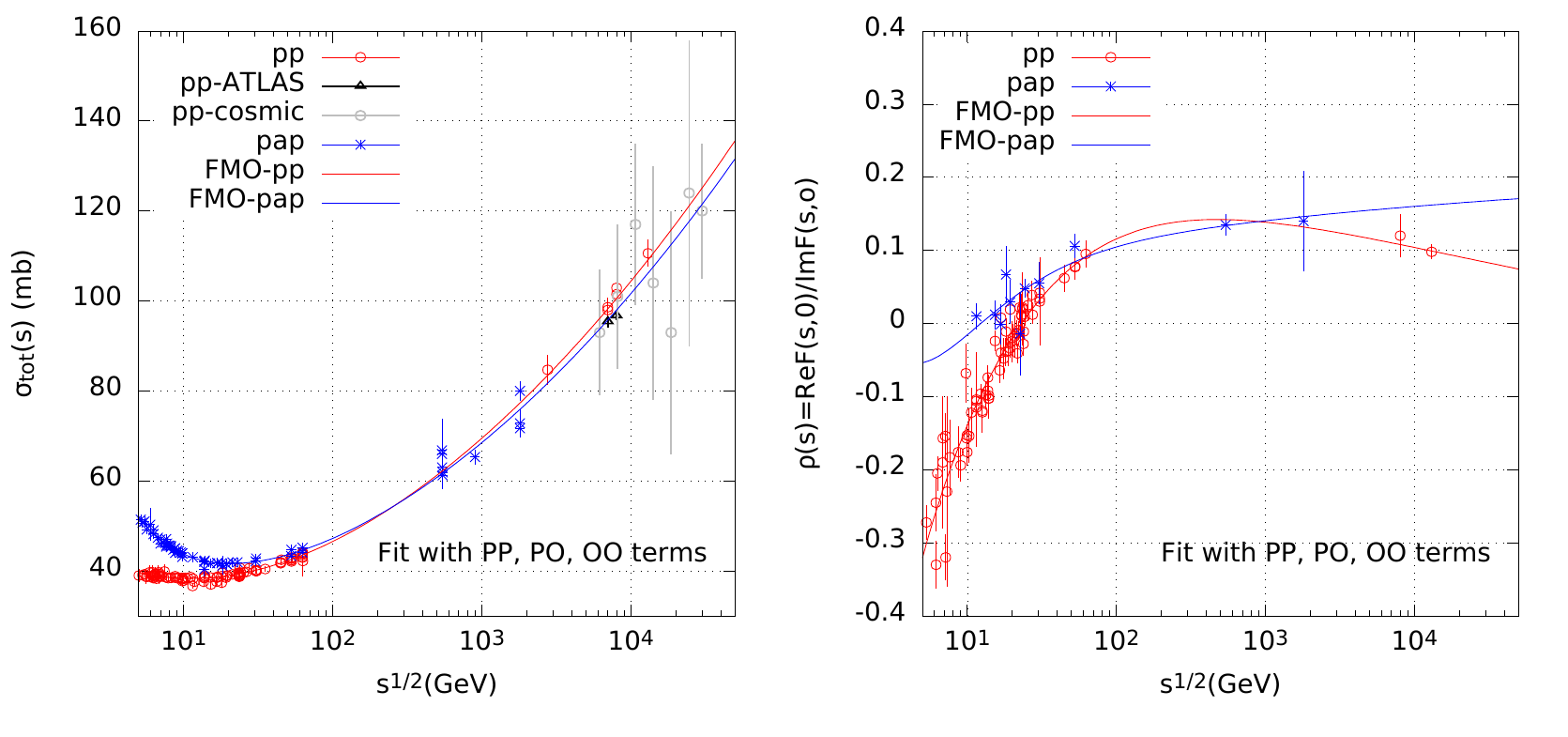}
	\caption{Total $pp$ and $\bar pp$ cross sections (left panel) and ratios $\rho^{pp}$ and $\rho^{\bar pp}$ (right panel) in the FMO model}
	\label{fig:2}
\end{figure}

\begin{figure}[H]
	\centering
	\includegraphics[width=0.45\linewidth]{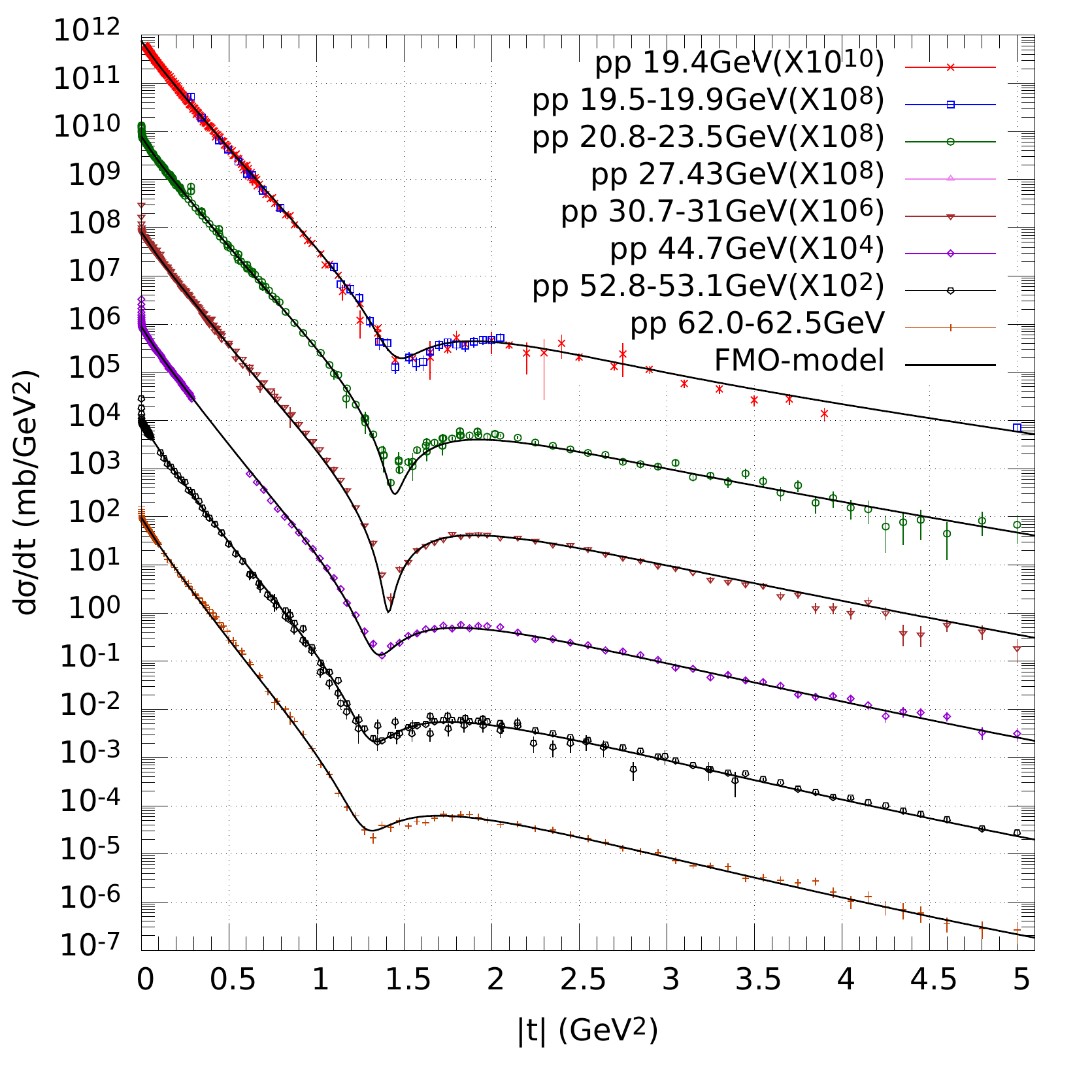}
	\includegraphics[width=0.45\linewidth]{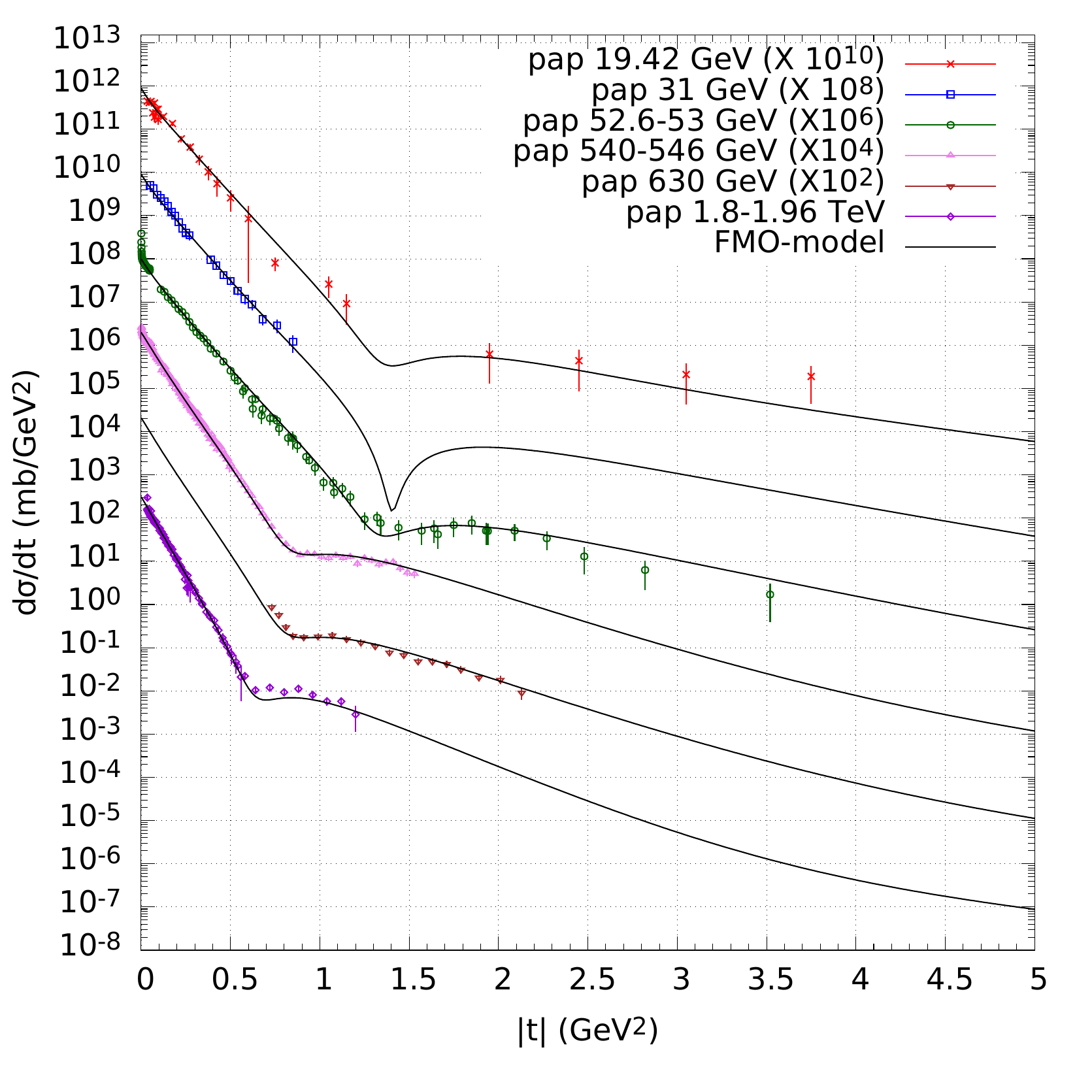}
	\caption{$pp$ (left panel)  and $\bar pp$ (right panel) differential cross sections at 19 GeV $<\sqrt{s}<$ 2 TeV  in FMO model}
	\label{fig:3}
\end{figure}

\begin{figure}[H]
	\centering
	\includegraphics[width=0.457\linewidth]{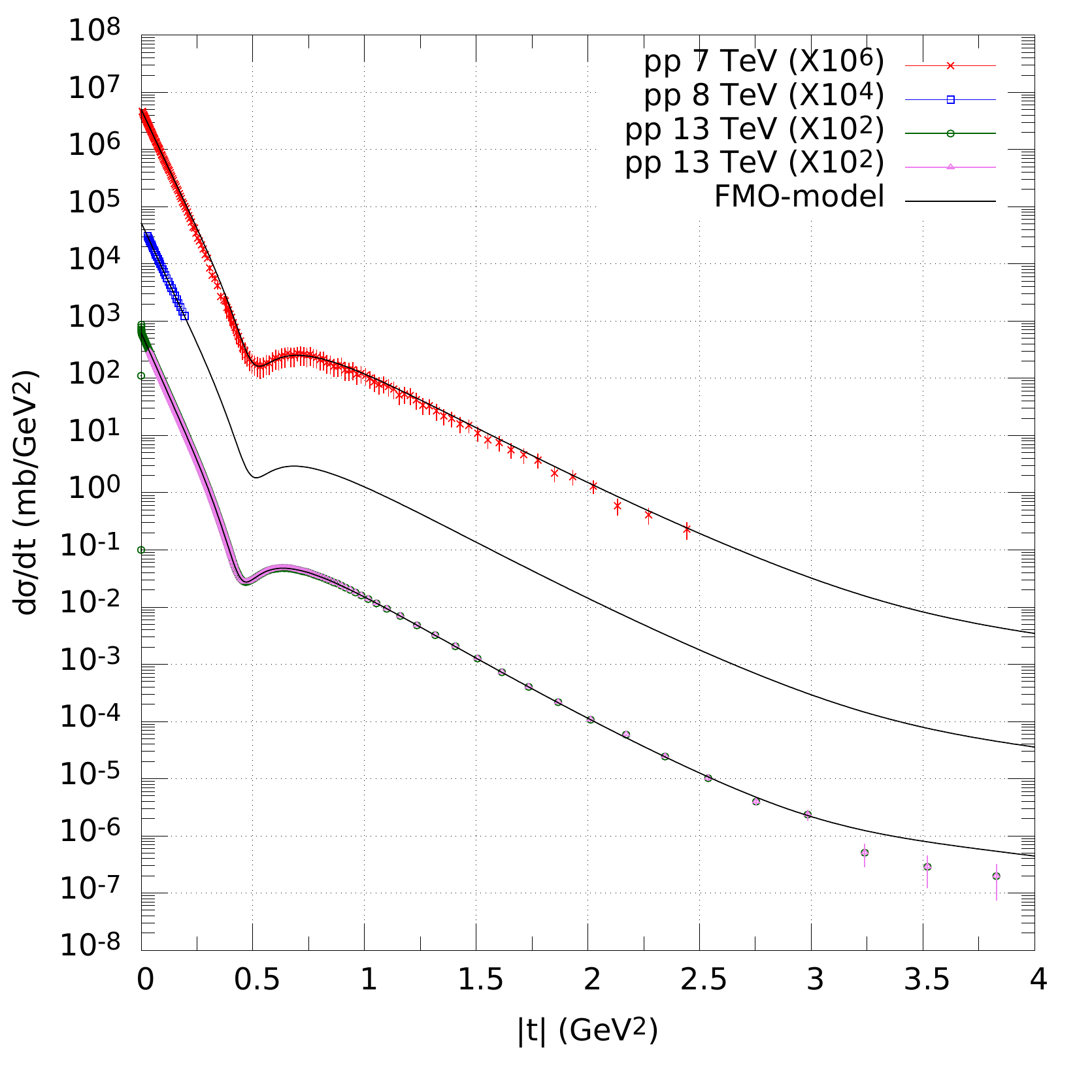}
	\caption{Differential $pp$ cross sections at LHC energies in FMO model}
	\label{fig:4}
\end{figure}

\section{Conclusion}
The TOTEM experiments firmly established the experimental existence of the Odderon, 45 years after its theoretical prediction. The Froissaron-Maximal Odderon (FMO) approach is the only existing model which describes the totality of experimental data (including the TOTEM results) in a wide range of energies and momentum transfers.

\bigskip

{\bf Acknowledgement}
	The authors thank Prof. S. Giani, Prof. K. Osterberg, Prof. T. Csörgö and Dr. J. Kaspar for their very useful comments.

\end{document}